# Ramifications and Diminution of Image Noise in Iris Recognition System

[1]Prajoy Podder, [2]A.H.M Shahariar Parvez, [3]Md. Mizanur Rahman, [4]Tanvir Zaman Khan

[1]Department of EEE, Ranada Prasad Shaha University, Narayanganj, Bangladesh.

[2,3]Department of CSE, Ranada Prasad Shaha University, Narayanganj, Bangladesh.

[1,4]Department of ECE, Khulna University of Engineering and Technology, Khulna-9203, Bangladesh.

[1]prajoypodder@gmail.com, [2]sha0131@gmail.com, [3]mizan173@gmail.com, [4]tzkhan19@gmil.com

*Abstract*—Human's Identity verification has always been an eye-catching goal in digital based security system. Authentication or identification systems developed using human characteristics such as face, finger print; hand geometry, iris, and voice are denoted as biometric systems. Among the various characteristics, Iris recognition trusts on the idiosyncratic human iris patterns to find out and corroborate the identity of a person. The image is normally contemplated as a gathering of information. Existence of noises in the input or processed image effects degradation in the image superiority. It should be paramount to restore original image from noises for attaining maximum amount of information from corrupted images. Noisy images in biometric identification system cannot give accurate identity. So Image related data or information tends to loss or damage. Images are affected by various sorts of noises. This paper mainly focuses on Salt and Pepper noise, Gaussian noise, Uniform noise, Speckle noise. Different filtering techniques can be adapted for noise diminution to develop the visual quality as well as understandability of images. In this paper, four types of noises have been undertaken and applied on some images. The filtering of these noises uses different types of filters like Mean, Median, Weiner, Gaussian filter etc. A relative interpretation is performed using four different categories of filter with finding the value of quality determined parameters like mean square error (MSE), peak signal to noise ratio (PSNR), average difference value (AD) and maximum difference value (MD).

## I. INTRODUCTION

Biometric using human iris deals with recognizing a person by his/her iris pattern extricated from the input iris image captured by camera. The prominent advantage of iris biometric system over other biometrics (Hand geometry, Retinal scan etc.) is that irises have vast pattern variability [1,2]. John Daugman claims that one conditional false reject probability occurs among 4 billion iris patterns [5]. More than 200 features of the iris can be analyzed such as furrows, coronas, stripes, freckles, zigzag collarette area etc [4]. Iris biometric is stable and remains the same for a person's lifetime compared to Face recognition, Palm recognition [2]. The use of contact lenses, glasses and even eye surgery cannot affect the iris characteristics [3]. But in the segmentation part of the iris recognition system, the Daugman's integrodifferential operator or Hough circular transform not always segments the iris appropriately when the input iris image contains noise in it.

Unwanted information mainly exists in noise signal, which deteriorates the image quality. It is delineated as a process which has an impact on the acquired image. Noise is introduced into images ordinarily at the time of transferring one stage to another stage and acquiring them. Images are affected by various types of noise, such as Salt-and-Pepper noise, Exponential noise, Gaussian noise, Speckle noise, Poisson noise, Rayleigh noise, Erlang (Gamma) noise, Periodic noise, etc. The main type of noise which is occurring during the image acquisition is called Gaussian noise. On the other hand, salt and pepper noise can occur in an unsecure communication channel during the transmission of image data. So, it is required to eliminate the noise from the image to get accurate data and devolve the visual quality of the images. There are various type of filters use for image noise reduction, such as Geometric and harmonic mean filter, wavelet based filtering, Median filter, bilateral filter, Alpha-trimmed mean filter, Adaptive filter, Notch filter, Weiner filter, Band pass filter, Band reject filter, SRAD filter etc. In order to remove noise and improve visual quality in this framework, some filtering algorithms have been applied and assessed using some performance parameters. The shape of the histogram of the images has been indicated the nature of the noise present in the images. Fig.1 shows a simplified block diagram of degradation and restoration scheme where I(x,y) is the input and degraded noisy image $G(x,y) = Degradation\ Function,\ H(x,y) * I(x,y) + Noise,\ N(x,y)$. Image obtained after the restoration block is called the filtered image [5].

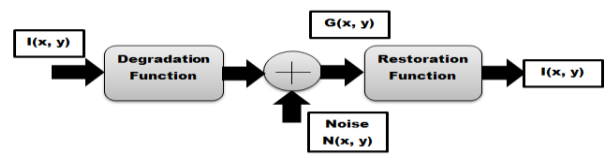

Fig. 1. Image degradation and restoration model

The research work can be structured as follows: the synopsis of the working process are described in second section, segment III represents different types of noises and their corresponding histograms, which are further treated by utilizing the noise removal filters. Fourth Section elucidates the Mean, Gaussian, Median and Weiner filters. Simulation results and analysis are described in part V, where mean square error, peak signal to noise ratio, average difference value, maximum difference value are determined precisely and the final passage concludes the paper.

## II. SYNOPSIS OF THE WORKING PROCESS

1. Transformation of the original input image that has been captured by camera or sensor to the gray scale images.

2. Different types of noise like Gaussian, speckle, Impulse noise, Uniform noise etc. have been added in the gray scale images possessing noise concentration (for Salt and Pepper noise), mean and variance (for Gaussian and speckle noise).

3. The noise contained images have been examined.

4. In order to reduce or diminish the unwanted noise from the images Median, Gaussian, Mean, Weiner filtering algorithm have been applied.

5. The Filter's efficiency have been observed by calculating mean square error (MSE), peak signal to noise ratio (PSNR), average difference value (AD) and maximum difference value.

6. Then the original input images in gray scale form and the noisy images have been included in the iris dataset to perform the matching process based on the wavelet and hamming distance process.

## III. TYPES OF IMAGE NOISE

### A. Salt and Pepper Noise

Salt corresponds to the maximum gray value (white). On the other hand Pepper resembles to the maximum gray value (Black). The usual value of pepper and salt noise is zero and 255 respectively for an eight bit image. Sharp and abrupt disturbances in the image can be the one of the main causes to create this noise. This noise may be called as Impulse noise. It can also be entitled as spike noise. An impulse noise enclosed image has dark pixels in bright sections and bright pixels in dark sections [8]. White and black points are appeared in digital gray scale images because of Salt-and-Pepper noise which are chaotically scattered along the image area (Refer to Fig.2). The Probability density function of (bipolar) impulse noise can be expressed by the following equation.

$$F(s) = \begin{cases} F_a, when\ s = a; \\ F_b, when\ s = b; \\ 0, otherwise \end{cases} \quad (1)$$

Gray level b will be appeared in the image as a light dot in the under the condition of b>a. Conversely, level a will appear like a dark dot. The impulse noise is called uni-polar when $F_a$ or $F_b$ is zero [1]. A median filter is skillful of mitigating salt and pepper noise because when the median value is taken by the pixel values that are varied from their neighboring pixels are replaced by a pixel value equal to the neighboring pixel value.

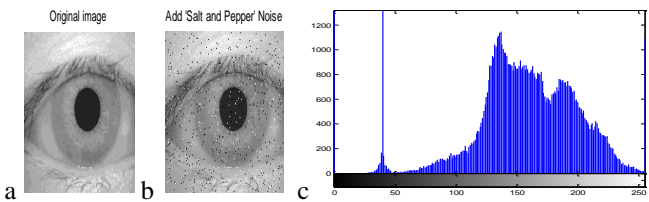

Fig. 2. Salt and Pepper noise mixed with original image & the corresponding histogram

### B. Gaussian Noise

Gaussian noise is such type of noise in which at each pixel position (i,j), the random noise value affecting the true pixel value is drawn from a Gaussian probability density function (PDF) (refer to fig.3(c)).If the random number of the noise field is added to the true value of the pixel, the noise is additive. It is created from internal electronic equipment of a circuit board, sensor noise due to high temperature or low light condition at the moment of image acquisition [10][11]. If the noise parameters possess same value for all pixels, the noise can be called homogenous Gaussian noise. If Mean value and SD value are the same for all pixels (i,j), the this can be denoted as Homogenous noise.

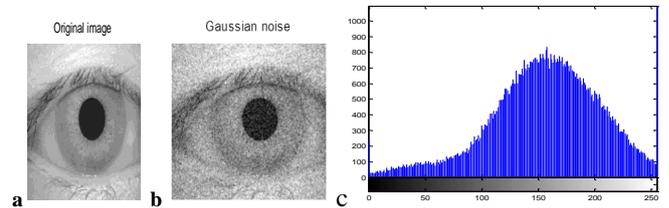

Fig. 3. Gaussian noise mixed with original image & the corresponding histogram

The PDF of Gaussian noise (fig. 4) is given by,

$$F(z) = \frac{1}{\sqrt{2\pi}\sigma} e^{\frac{-(z-\mu)^2}{2\sigma^2}} \quad (2)$$

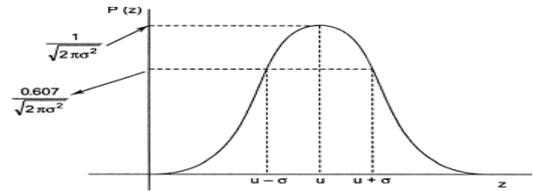

Fig. 4. PDF of Gaussian noise

### C. Speckle Noise

Speckle noise is usually found in all coherent imaging systems (complex amplitude linear imaging operation) such as laser light, acoustic, SAR and medical ultrasound imagery. This noise follows a gamma distribution which can be expressed as,

$$F(g) = \frac{g^{\alpha-1}}{(\alpha-1)!a^\alpha} e^{\frac{-g}{a}} \quad (3)$$

Where, g means the gray level and $\alpha$ is the variance. Speckle has the unfortunate characteristic of decreasing into the high sensitivity region from human vision to spatial frequency. It can limit the effective application (edge detection) of automated computer analysis algorithms.

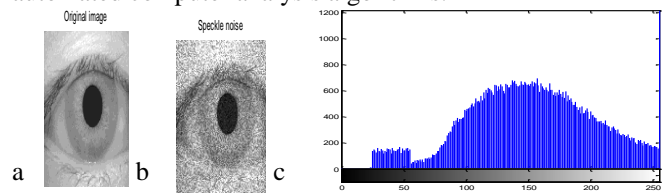

Fig. 5. Speckle noise mixed with original image & the corresponding histogram

Speckle noise is considered as a primary factor for limiting the contrast resolution in diagnostic medical imaging as well as limiting the detectability of small low contrast regions. Speckle noise has multiplicative properties for images having low contrast. Complex amplitude of Speckle noise (Jain 2006) [19] is,

$$a(x,y) = a_R(x,y) + ja_I(x,y) \quad (4)$$

Intensity, $s(x,y) = |a(x,y)|^2 = a_R^2 + a_I^2$

Speckle Index, $SI = \frac{1}{mn}\sum_{i=1}^{m}\sum_{j=1}^{n}\frac{\sigma(i,j)}{\mu(i,j)}$

*D. Uniform Noise*

Uniform noise is induced by quantizing the pixels of a sensed image to a number of distinct levels. In this noise, the level of the gray values is uniformly distributed across a specified range [9].

The PDF of uniform noise (fig. 6) is given by,

$$P(z) = \begin{cases} \frac{1}{b-a}, & if\ a \leq z \leq b \\ 0, & otherwise \end{cases} \quad (5)$$

Mean value is given by, $\mu = \frac{a+b}{2}$

Variance is given by, $\sigma^2 = \frac{(b-a)^2}{12}$

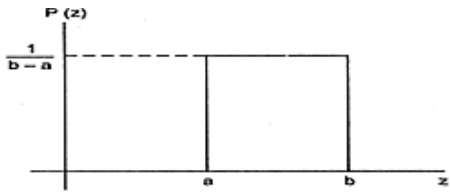

Fig. 6. PDF of Uniform noise

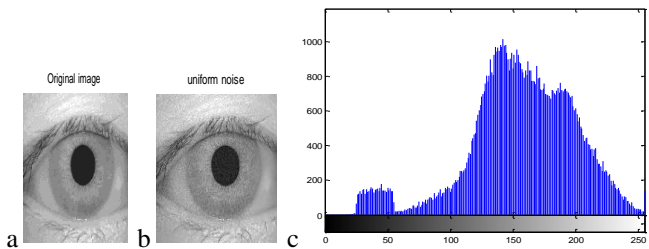

a b c

Fig. 7. Image with Uniform noise & the corresponding histogram

Fig. (2,3,5,7) describes sequentially the Salt-and-Pepper noise, Gaussian noise, Speckle noise, unifrom noisy images and their corresponding histrogrames.

## IV. IMAGE DE-NOISING FILTERS

Image de-noising is a prominent image pre-processing stage in sophisticated imaging applications such as bio medical image investigation, signal (audio, compressed video) analysis, Satellite and radar signal analysis etc. Each application has its special necessities. Since, noise elimination of medical signals such as ECG, EEG, EKG etc. is very sensitive, proper de-noising filtering algorithm must be applied. Since, noise can be mixed with original medical data or image during transmission through telemedicine. Noise free image is essentially needed for line, point and edge detection on an image. Smoothing filters and sharpening filters can be considered as two main categories at the time of eliminating or reducing image noise.

High frequency components can be contained in the image. Smoothing filters can play an important role for suppressing those high frequencies. On the other hand, sharpening filters are used to extract and highlight fine details from an image and also to enhance some blurred details. It can annihilate the low frequencies in the image. Process of restoring an image from the noisy image without least loss of information can be classified not only spatial domain but also frequency domain categories. This grouping is centered on modifying the Fourier transforms of an image. Noise removal is complex in the frequency domain as compared to the spatial domain. The spatial domain noise removal requires very less processing time. Noise removal algorithms should provide a satisfactory amount of noise removal and also help preserve the edges. There are two types of filters which can satisfy the stated conditions: linear and non-linear with their significant advantages and disadvantages. The faster processing linear filters are failed to preserve edges. On the contrary the slower processing non-linear filters are successful of conserving edges. These noise removal filters have been used in this work for performance evaluation. Now, the working process of the four filters (mean, median, Gaussian and Weiner filter) with necessary equations has been discussed below for the purpose of diminishing the noise that affected the fresh input image.

*A. Median Filter*

The effective solution for reducing the amount of Salt and Pepper noise is to deployment of a median filter. Because of this, the pixel's intensity in the image may be accidentally increased. A best solution to eliminate the mentioned problem is to find out the median of the grey value in the neighborhood pixel and then the extreme value is replaced by the median value [10]. The mathematical expression of median filter is given by the following equation:

$$f'(x,y) = \underset{(i,j)\in S_{xy}}{median}\{g(i,j)\} \quad (6)$$

Generally a 3x3 window or a 5x5 window is chosen and then determining the center of the pixel [i,j], the intensity of the that pixel is calculated for each window. But after applying this filter, the Gaussian noise of smaller amplitude still remains [12]. Median filter can be separable or non-separable.

Fig. 8 demonstrates the computation of the median filter of size 3x3 pixels. The nine pixel values extracted from the 3x3 image region are arranged as a vector that is sorted and the resulting center value is taken as the median.

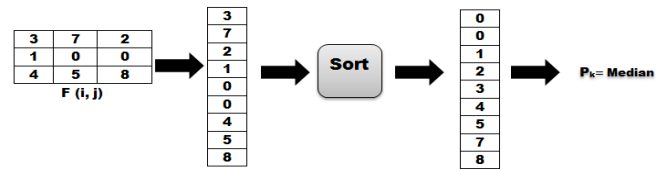

Fig. 8. Estimation of a F(3,3) pixel median filter

Median filters are primarily effective in the existence of not only bipolar but also unipolar impulse noise [1].

*B. Gaussian Filter*

Gaussian filter is a filter which exploring the relationships between the spatial and frequency domain. The taper of this filter reduces ringing but it also reduces the degree of smoothing [13] [14]. For a filter mask of nxn size, coefficients of Gaussian filter are obtained by the equation,

$$h_G(x,y) = e^{-\frac{1}{2}\left(\frac{d}{\sigma}\right)^2} \text{ where } d = \sqrt{x^2 + y^2} \qquad (7)$$

Impulse response of 1D Gaussian filter can be mathematically expressed as,

$$G(x) = \sqrt{\frac{a}{\pi}} e^{-ax^2} \qquad (8)$$

Considering standard deviation as a parameter the above equation can be written in eqn. (8). Gaussian filter becomes an uniform filter when $\sigma$ becomes infinite.

$$G(x) = \frac{1}{\sqrt{2\pi}.\sigma} e^{-\frac{x^2}{2\sigma^2}} \qquad (9)$$

*C. Mean Filter*

Mean filtering can be also a substitute approach in order to suppress Gaussian noise as well as speckle noise [10]. Arithmetic mean filter can be expressed mathematically by the following equation.

$$a(i,j) = \frac{1}{N} \sum_{n=i-1}^{i+1} \sum_{n=j-1}^{j+1} g(m,n) \qquad (10)$$

Here N and g(m,n) represent total number of pixels in the neighborhood window and the pixel value in the neighborhood respectively.

*D. Weiner Filter*

Wiener filter can be used to diminish the portion of noise exist in the image matrix when it is compared with an estimation of the desired pure noise free signal. The transfer function of this filter is chosen to minimize MSE using statistical information on not only image but also noise fields [15].The main drawback of this filter is that to get an optimum solution; one should know the statistics of the image and noise.

Image recovery using noise adaptive Weiner filtering is given by [16],

$$f'(x,y) = m_f(x,y) + \frac{\sigma_f^2(x,y)}{\sigma_f^2(x,y)+\sigma_n^2(x,y)} (g(x,y) - m_f(x,y)) \qquad (11)$$

V. EXPERIMENTAL RESULTS WITH ANALYSIS

With the help of Matrix laboratory software (MATLAB) R2014 version the evaluation of the noise removal filters has been examined on four different types of noises.

Fig. (9,10,11,12) describes the ramifications or consequences of filtering sequentially in order to reducing or diminishing the salt and pepper noise, gaussian noise, speckle noise and unifrom noise in CASIA database version 1.0. Fig. (9(i), 10(i), 11(i), 12(i)) shows the the orginal images, fig. (9(ii), 10(ii), 11(ii), 12(ii)) shows thew simulation result of sequentially salt and pepper noise, gaussian noise, speckle noise and unifrom noise images. fig. (9(iii), 10(iii), 11(iii), 12(iii)) represents median filter for eliminating the salt and pepper noise, gaussian noise, speckle noise and uniform noise from the images; fig. (9(iv,v,vi), 10(iv,v,vi), 11(iv,v,vi), 12(iv,v,vi)) represents sequentially mean, gaussian and winer filters for diminishing or eliminating the salt and pepper noise, gaussian noise, speckle noise and unifrom noise from the degraded (noisy) images.

*A. CASIA Iris Database V1.0*

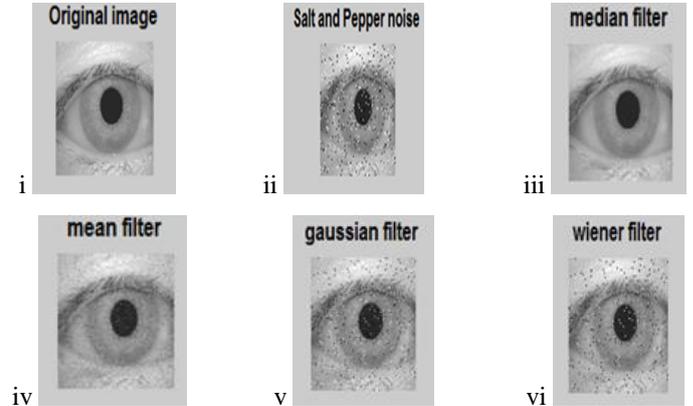

Fig. 9. Ramifications of filtering for removing Salt and Pepper noise with respect to original image.

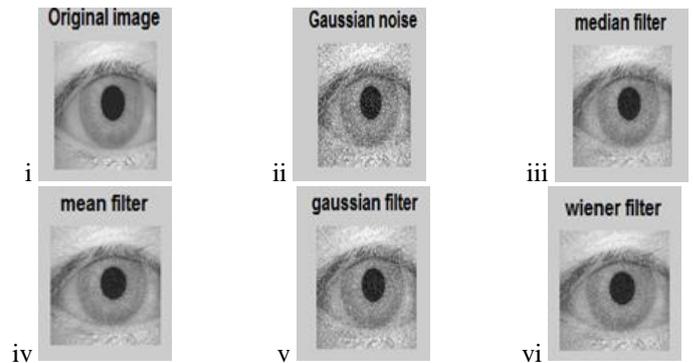

Fig. 10. Ramifications of filtering for eliminating Gaussian noise with respect to original eye image.

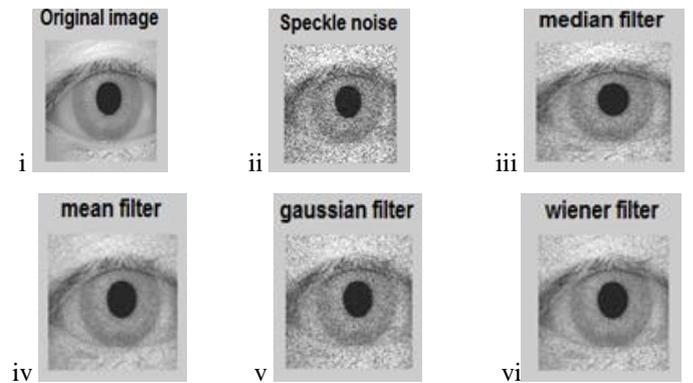

Fig. 11. Effect of filtering for eliminating Speckle noise

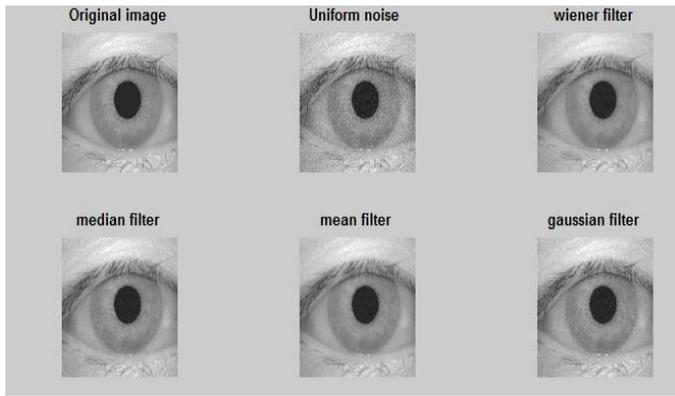

Fig. 12. Consequences of filtering for removing Uniform noise

Fig. (13,14,15,16) describes the effect of filtering sequentially for removing the salt and pepper noise, gaussian noise, speckle noise and unifrom noise in UBIRIS database. Where, median, mean, gaussian and winer filters are used for removing the salt and pepper noise, gaussian noise, speckle noise and unifrom noise from the images.

*B. UBIRIS Database*

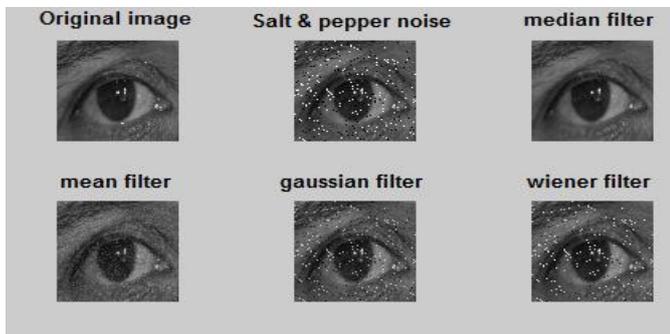

Fig. 13. Consequence of filtering for eliminating Salt and Pepper noise

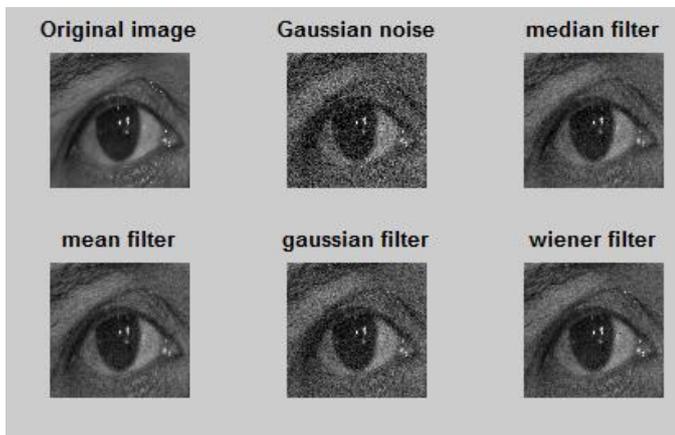

Fig. 14. Consequence of filtering for removing Gaussian noise

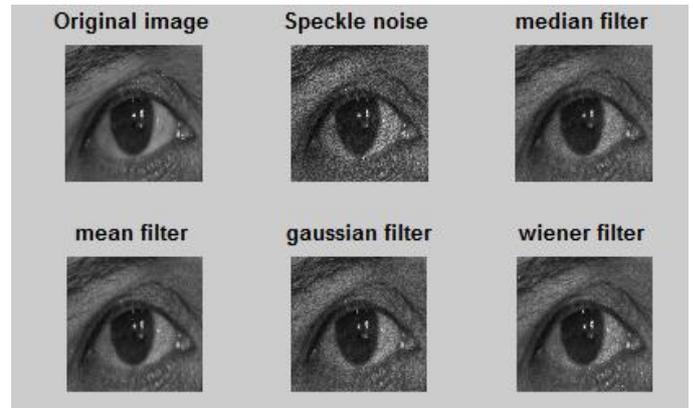

Fig. 15. Effect of filtering for eliminating Speckle noise

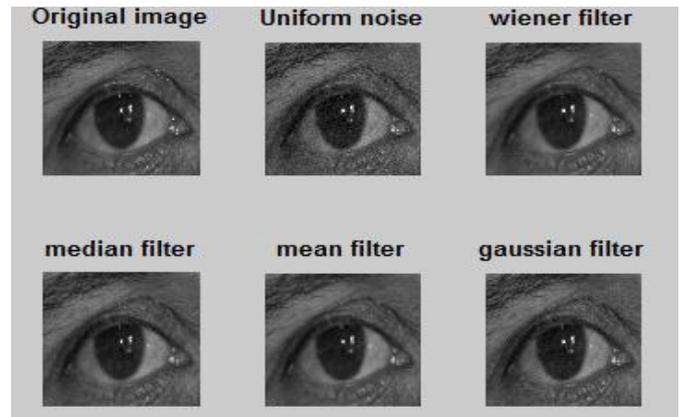

Fig. 16. Consequence of filtering for abolishing Uniform noise

*C. Mean Square Error*

Mean square error can be calculated by averaging the square of the errors between the recovered noises eliminated image and input noise image [18]. Different image algorithms can be compared by MSE.

$$MSE = \frac{1}{MN}\sum_{x=1}^{M}\sum_{y=1}^{N}[f(x,y) - f'(x,y)]^2 \qquad (12)$$

Here, MxN indicates the image pixel quantity, $f'(x,y)$ is the corrupted (noisy) image and $f(x,y)$ is the filtered (noise free) image. The lower MSE provides best strained image. A lower amount of MSE indicates lesser error in the reconstructed image.

TABLE I. MSE VALUES FOR VARIOUS DE-NOISING FILTERS AND NOISES

| Image De-noising Filter | Types of noise | | | |
|---|---|---|---|---|
| | Salt and pepper | Gaussian | Uniform | Speckle |
| Mean | 50.87 | 87.82 | 35.56 | 101.13 |
| Median | 15.81 | 78.13 | 31.38 | 87.28 |
| Gaussian | 25.78 | 46.46 | 9.75 | 66.83 |
| Weiner | 22.31 | 81.70 | 28.63 | 97.11 |

Median filter can provide approximately better result in eradicating salt-and-pepper noise. If impulse noise is large then median filter gives poor result. Though Weiner filter is more susceptible to noise, it has good performance for lessening noise [20]. The Performance of Weiner filter is good for removing Gaussian and Poisson noise. Gaussian filter gives comparatively less MSE in speckle noise. Table I shows the MSE values for different filters and noises of CASIA iris image (Fig.9-Fig.12). In the case of Salt and Pepper noise, the MSE of median filtered image is 15.81 which are smaller than Gaussian, Mean and Weiner filter.

*D. Peak Signal to noise ratio*

The Peak Signal to Noise Ratio (PSNR) can be calculated by the following equation,

$$PSNR = 10 \log_{10}\left[\frac{255 \times 255}{MSE}\right] \quad (13)$$

It normally reflects the quality of the restored image. A image is called best quality for high value of PSNR of a particular noise case [21]. Median filter produces higher PSNR values for removal of Salt and pepper noise where Gaussian filter provides large value of PSNR in order to remove Gaussian and Uniform noise .Table II shows the PSNR values in dB for different filters and noises of CASIA iris image (Fig.9-Fig.12).

TABLE II. PSNR VALUES FOR VARIOUS TYPES OF NOISE

| Filter applied | Categories of noise | | | |
|---|---|---|---|---|
| | Salt and pepper | Gaussian | Uniform | Speckle |
| Mean | 31.10 | 28.73 | 32.66 | 28.12 |
| Median | 36.17 | 29.24 | 33.19 | 28.76 |
| Gaussian | 34.05 | 31.49 | 38.28 | 29.91 |
| Weiner | 34.68 | 29.04 | 33.59 | 28.29 |

*E. Average Difference (AD)value*

The average difference can be derived normally by averaging root square of the errors between the restored image and noisy image.

$$AD = \frac{1}{MN}\sum_{x=1}^{M}\sum_{y=1}^{N}[f(x,y) - f'(x,y)] \quad (14)$$

Where, M and N denote the row and column dimensions of the image respectively. A cleaner image normally has a lower amount of average difference value [4]. In the case of salt-and-pepper noise, average difference value of the median filter is less and due to this reason it may be called an ideal filter for such noise. Performance of Median and Weiner filter is comparatively good for Gaussian, speckle and Poisson noise. Table III shows the AD values for different filters and noises of CASIA iris image have been showed Table III (Fig.9-Fig.12).

TABLE III. AD VALUES FOR VARIOUS FILTERS AND NOISES

| Filter implemented | Types of noise | | | |
|---|---|---|---|---|
| | Salt and pepper | Gaussian | Uniform | Speckle |
| Mean | 6.99 | 10.36 | 4.10 | 14.20 |
| Median | 3.85 | 9.37 | 3.14 | 13.06 |
| Gaussian | 2.88 | 4.45 | 1.62 | 6.11 |
| Weiner | 3.45 | 8.61 | 3.05 | 11.66 |

*F. Maximum Difference (MD)value*

It is the maximum absolute difference value of the recovered image and noisy image.

$$MD = Max(|f(x,y) - f'(x,y)|) \quad (15)$$

Table IV shows the MD values for different filters and noises of CASIA iris image .The picture quality is poor if the value of Maximum difference is large. The value of median filter is comparatively less than other mentioned filters.

TABLE IV. Maximum difference (MD) values for Several filters and noises

| Image De-noising Filter | Types of noise | | | |
|---|---|---|---|---|
| | Salt and pepper | Gaussian | Uniform | Speckle |
| Mean | 252 | 251 | 253 | 254 |
| Median | 225 | 215 | 205 | 195 |
| Gaussian | 254 | 254 | 254 | 255 |
| Weiner | 252 | 251 | 253 | 254 |

*G. Effect of Hamming distance*

Fig.17and fig.18 represents the hamming distance of noise free and noisy CASIA database iris images. It can be observed that, noisy images have greater hamming distance value than the noise free image.

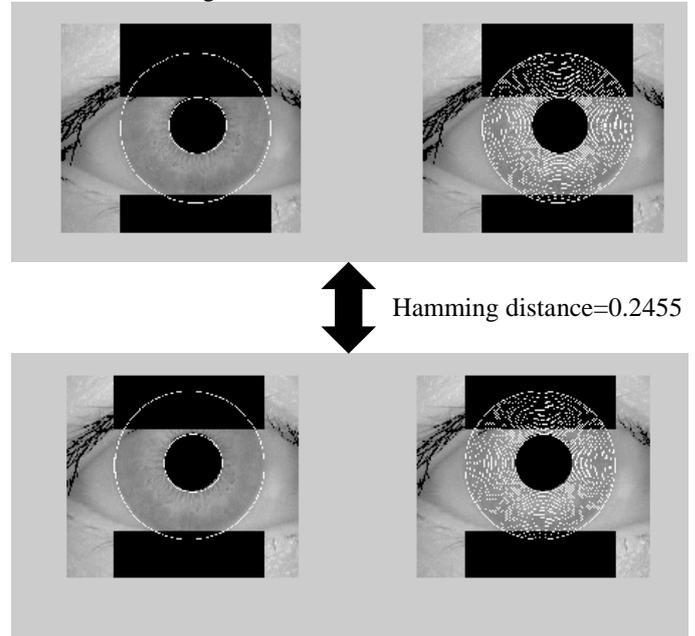

Hamming distance=0.2455

Fig. 17. Hamming distance of noise free CASIA database iris images

As a result, it affects the Iris recognition system. Because of larger hamming distance, the same individual person's iris cannot be matched. Noisy image also changes the iris template or mask. So it should be bear in mind that amount of image noise must be negligible so that deviation of hamming distance value does not occur and local binary pattern of normalized iris template remains noise free. The higher amount of diminishing noise gets the least probability of false acceptance rate and false rejection rate of the iris recognition method.

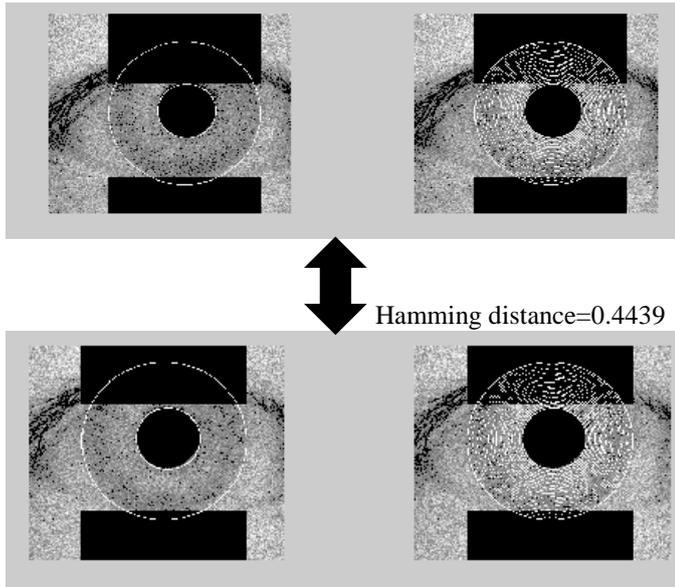

Hamming distance=0.4439

Fig. 18. Hamming distance of noisy CASIA database iris images

## VI. CONCLUSION

Iris recognition system is considered as one of the best biometric technology because of its unique characteristics. But Iris recognition system can be affected due to noise and it affects the segmentation process of the iris as well as the feature extraction and matching process. So.it should be very careful that noises do not contaminate the iris image and if noise added to the image, noise must be removed by applying the various de-noising filters. At the time of eliminating noise, it should also be careful that Eyelids, eyelashes or any part of the iris region are not considered as noisy information or are not filtered by de-noising filters. Four types of noise and basic noise removal filters like mean, median etc. in order to remove those noises have been discussed in the paper. The ramifications i.e. effects of noisy images in the feature extraction and matching section (with respect to the hamming distance) has also been observed practically.

## REFERENCES

[1] Jaakko Astola, Pauli Kuosmanen, "Fundamentals of Nonlinear Digital Filtering", CRC Press, 1997.

[2] John Daugman, "Recognizing persons by their iris patterns", Cambridge University, Cambridge, UK., 2001.

[3] Prem Kalra, "Computer Vision, Graphics and Image Processing: 5th Indian Conference (ICVGIP), Madurai, India, December 13-16, 2006, Proceedings", Springer Science & Business Media, 2006.

[4] Prajoy Podder, Tanvir Zaman Khan, Mamdudul Haque Khan, M. Muktadir Rahman, Rafi Ahmed, Md. Saifur Rahman, "An efficient iris segmentation model based on eyelids and eyelashes detection in iris recognition system", International Conference on Computer Communication and Informatics (ICCCI),India, 2015.

[5] Charles Boncelet, "Image Noise Models" in Alan C.Bovik, Handbook of Image and Video Processing, 2005.

[6] Gajanand Gupta, "Algorithm for Image Processing Using Improved Median Filter and Comparison of Mean, Median and Improved Median Filter" (IJSCE) ISSN: 2231-2307, Volume-1, Issue-5, November 2011.

[7] Christos P. Loizou, Constantinos S. Pattichis, "Despeckle Filtering Algorithms and Software for Ultrasound Imaging", Morgan & Claypool Publishers, 2008.

[8] Narasimharao, B. Pandu Ranga,, "Handbook of Research on Science Education and University Outreach as a Tool for Regional Development", IGI Global, 2017.

[9] J.M Li et al, "Image based Fractal Description of microstructures", Springer.

[10] Maria Petrou, Costas Petrou, "Image Processing: The Fundamentals", John Wiley and Sons Publishers, 2010.

[11] Raymond H. Chan, Chung-Wa Ho, and Mila Nikolova, "Salt-and-Pepper Noise Removal by Median-Type Noise Detectors and Detail-Preserving Regularization", IEEE Transactions on Image Processing Volume: 14, No: 10, pp. 1479 – 1485, November 2005.

[12] Charles Boncelet,"Image Noise Models", In Alan C. Bovik. Handbook of Image and Video Processing. Academic Press. ISBN 0-12-119792-1,2005.

[13] Abhishak Yadav, Poonam Yadav, "Digital Image Processing", Jan 1, 2009.

[14] Jayaraman, "Digital Image Processing", Tata McGraw-Hill Education, 2011.

[15] Oge Marques, "Practical Image and Video Processing Using MATLAB", John Wiley & Sons, Jul 28, 2011.

[16] Madhu S. Nair, K. Revathy, and Rao Tatavarti, "Removal of Salt-and-Pepper Noise in Images: A New Decision-Based Algorithm", Proceedings of the International MultiConference of Engineers and Computer Scientists,Vol I 19-21 March, 2008, Hong Kong.

[17] Muthuselvi.s and Narmadha.d., "Survey on Removal of Universal Noise from Images", International Journal of Computer Applications ,Volume: 58, Number:7,pp.27-31, November 2012.

[18] Aparna Vyas, Soohwan Yu, Joonki Paik, "Multiscale Transforms with Application to Image Processing", Springer, 2017.

[19] Anil K. Jain, "Fundamentals of Digital Image Processing", 4[th] Edition, Prentice Hall, 1989.

[20] Tinku Acharya, Ajoy K. ray, "Image Processing: Principles and Applications", John Wiley and Sons Publisher, 2005.

[21] Prajoy Podder and Md. Mehedi hasan, "A Meta Study of Reduction of Speckle Noise Adopting Different Filtering Techniques", 3rd International Conference on Electrical Engineering and Information Communication Technology (ICEEICT), MIST, 2016.